\documentclass[12pt,reqno]{article}
\pdfoutput=1
\usepackage{titlesec}
\usepackage[perpage]{footmisc}
\titleformat{\paragraph}
{\normalfont\normalsize\bfseries}{\theparagraph}{1em}{}
\titlespacing*{\paragraph}
{0pt}{3.25ex plus 1ex minus .2ex}{1.5ex plus .2ex}
\usepackage{graphicx}	
\usepackage{soul}
\usepackage{amsmath,amssymb,latexsym}	
\usepackage[square,numbers]{natbib}
\usepackage[normalem]{ulem}
\usepackage{notoccite}

\usepackage[totalwidth=460truept,totalheight=600truept]{geometry}

\makeatletter%
\xdef\Equ@width{\the\@tempdima}

\usepackage[T1]{fontenc}
\usepackage{color}

\global\arraycolsep=1truept
\newcommand{\beq}{\begin{equation}}
\newcommand{\eeq}{\end{equation}}
\newcommand{\bea}{\begin{eqnarray}}
\newcommand{\eea}{\end{eqnarray}}

\newcommand{\cc}[1]{c_\mathrm{#1}}

\definecolor{darkred}{rgb}{.8,0,0}

\definecolor{darkblu}{rgb}{0,0,.8}

\begin{document}

\vskip 1.4truecm

\begin{center}

\pagestyle{empty}
\vspace{1.2cm}\end{center}
\begin{center}
\LARGE{\bf Entanglement Entropy at Large-N}\\[12mm] 

\vspace{9mm}

\large
\textsc{ P.~Talavera}

\vspace{8mm}

\footnotesize{
Department of Physics,
Polytechnic University of Catalonia,\\ Diagonal 647,
Barcelona, 08028, E\\
}


\vspace{4mm}
%
{\footnotesize\upshape\ttfamily ptalavera@protonmail.ch} \\

\vspace{8mm}
\noindent

\small{\bf Abstract} \\
\end{center}
\begin{center}
\begin{minipage}[h]{\textwidth}
I show that at early times the evaporation process for a stack of NS5-branes at high energy is suppressed in the large-N limit.
At much later times, the new saddles in the gravitational action are no longer suppressed at large-N, and evaporation proceeds as usual.
This effect is due to the non-invariance of the metric under coordinate parameterisation.
This fact introduces a parametric dependence in the thermodynamic quantities, which can lead to quantities that are  
suppressed in a particular corner of the parameter space. 

\end{minipage}
\end{center}
\newpage
\setcounter{page}{1}
\pagestyle{plain}

\section{Motivations}

In this short note, I would like to explore a novel behaviour that occurs 
in black holes associated with a stack of NS5-branes at the Hagedorn temperature. This behaviour is a direct consequence of a large-N limit\footnote{From now on, N will denote the number of NS5 branes.} and is rooted in the peculiarity of the system. 
The setup is quite appealing, first because Little String Theory (LST) is a non-local field theory without gravity, and one hopes to use this knowledge as a precursor to understanding general settings of string theory. And secondly, the thermodynamic variables indicate that the system is at the Hagedorn temperature, a point where the partition function is poorly defined.

The theoretical framework is defined on the world-volume of NS5-branes in the limit of vanishing string coupling, $g_\alpha \to 0\,.$ It was shown in \cite{Callan:1991at} that LST
is holographically dual to string theory on CGHS black hole. We shall rely heavily on this fact.

A Semi-classical analysis \cite{Lorente-Espin:2007jqj}, shows that the radiation associated with the
black hole inside the LST is purely thermal and therefore
indistinguishable from white noise \cite{10.1119/1.1773578}.
This fact has deeper implications \cite{Lorente-Espin:2007jqj}: The
probability of emitting a shell of energy $\omega_1 + \omega_2$ is equal to the probability of emitting two independent shells with the same total amount of energy, $\omega_1 + \omega_2$. As a direct consequence
the radiation always comes as a pure state, the Hilbert space can be factorized into two disjoint parts, 
\begin{equation}
\label{fact}
    H= H_\mathrm{in}\oplus  H_\mathrm{out}\,,
    \end{equation}
corresponding to states located at the inner and
outer sides of the event horizon respectively. Note that in \eqref{fact} is missing
an interaction piece $H_\mathrm{int}$ which is mandatory if an outside observer located at the spatial infinity wants to extract information about the system \cite{deco}.

The aim of this work is to clarify whether this pattern is due to the setup temperature or to some deeper cause. To do this, I look at the entanglement entropy in the spirit of \cite{Almheiri:2019yqk}. Our results are unambiguous: the large-N limit suppresses semi-classical configurations, and as a consequence the system does not interact with the environment. Corrections, i.e. non-trivial saddles of the Euclidean path integral
\cite{Penington:2019npb,Almheiri:2019psf}, are not suppressed in this limit and
are fully responsible for the unitary evaporation of the black hole. In the spirit of \cite{Almheiri:2019qdq} this new configurations 
will be complex solutions of the gravitational equations and
correspond to saddles relevant to the unitary Page curve. 

For those familiar with QCD, this effect has a close parallel in the decay $\rho\to \pi\pi$ at large-N$_c$. 
Inserting the hadronic spectrum of large–N$_c$ QCD as a proxy for the real hadronic spectrum provides rather good approximation.
In such a model the spectrum of the theory in the large–N$_c$ limit consists of an infinite number of \emph{narrow stable}
meson states that cannot decay because of the zero width \cite{Golterman:2001pj}. 

\section{NS5-branes and LST backgrounds}
\label{sec:LST}

Holography relates LST to string theory in the near-horizon geometry of NS5-branes
\cite{ABKS}. Our starting point is the supergravity solution for N coincident near-extremal NS5-branes in the string frame \cite{giddingsstrominger}
\begin{align}
\label{ns5}
ds^2&=-\left(1-\frac{r_0^2}{r^2}\right) dx_1^2+\sum_{j=2}^6 dx_j^2+
\left(1+\frac{N}{m_s^2 r^2}\right) \left(\frac{dr^2}{1-r_0^2/r^2}+r^2 d\Omega_3^2\right)\,,
\\ \nonumber
e^{2\phi}&=g_s^2 \left(1+\frac{N}{m_s^2 r^2}\right)\,,
\end{align}
where $r_0$ is the location of the horizon, $g_s$ is the asymptotic string coupling constant and $m_s^2$ is essentially the string tension. The index $i=2\,,\cdots\,,6$ corresponds to the flat directions along the five-brane, and
$d\Omega_3$ is the line element of the unit 3 sphere.
The thermodynamics of the NS5-brane is 
\begin{align}
\label{thns5}
\frac{E_\mathrm{NS5}}{V_5}&=\frac{1}{(2\pi)^5 (\alpha^\prime)^3}\left(\frac{N}{g_s^2}+\frac{r_0^2}{g_s^2\alpha^\prime}\right)\,,\\
\beta_\mathrm{NS5}&=2\pi \sqrt{\frac{N}{m_s^2}}\sqrt{1+\frac{m_s^2 r_0^2}{N}}\,.
\end{align}
The first term, proportional to $\frac{N}{g_s^2}$, in \eqref{thns5} is the tension between the extremal NS5-branes.
There are several limits that can be run on \eqref{ns5}. One of them is a high energy limit, for which we approach the near horizon 
in the decoupling limit
\begin{equation}
\label{limith}
r_0\rightarrow 0\,,\quad g_s\rightarrow 0\,,\quad \frac{r_0^2}{g_s^2\alpha^\prime}\equiv\mathrm{fixed}\,.
\end{equation}
The resulting theory is conjectured to be dual to a Little String Theory \cite{ABKS}.
To take the limit, it is more convenient to change the variables to
\begin{equation}
u:=\frac{r}{g_s l_s^2}\,,\quad u_0:=\frac{r_0}{g_s l_s^2}\,,
\end{equation}
after which \eqref{ns5} becomes
\begin{align}
\label{lst}
ds^2&=-\left(1-\frac{u_0^2}{u^2}\right) dx_1^2+\sum_{j=2}^6 dx_j^2+
\frac{N}{m_s^2 u^2}\left(\frac{du^2}{1-u_0^2/u^2}+ u^2 d\Omega_3^2\right)\,,\quad
    e^{2\phi}=N\frac{m_s^2}{u^2}\,.
 \end{align}
The change \eqref{limith} is motivated as follows: in Type IIB string theory $u$ corresponds to the mass of a string stretching between two 
NS5-branes, while in Type IIA $u l_s^{-1}$ is the string tension of an open D2-brane stretched  between two D5-branes.
It is the thermodynamics associated with this setup \eqref{lst} \cite{Cotrone:2007qa} that will be more relevant to our discussion below 
\begin{align}
\label{thlst}
\frac{E_\mathrm{LST}}{V_5}=\frac{1}{(2\pi)^5 (\alpha^\prime)^3}\left(\frac{N}{g_s^2}+u_0^2 l_s^2\right)\,,\quad
\beta_\mathrm{LST}=2\pi \sqrt{\frac{N}{m_s^2}}\,,\quad \frac{S_\mathrm{LST}}{V_5}=\frac{1}{(2\pi)^4 (\alpha^\prime)^2}\sqrt{N} u_0^2 l_s\,.\quad 
\end{align}
Note that the temperature is independent of the energy. This fact allows to tune
independently both, energy and temperature, and makes the free energy to vanish.
It is clear, from \eqref{thlst}, that if $u_0^2 l_s^2 \gg N/g_s^2$ follows that $S=\beta_\mathrm{LST}E_\mathrm{LST}$ which is the leading behaviour for the entropy of a gas of weakly interacting closed strings at the Hagedorn temperature, the latter identified with that in \eqref{thlst} \cite{SALOMONSON1986349}. 

Since we are interested in describing the time evolution across the horizon, which is necessary to study the two-sided correlation functions, we shall adopt Kruskal coordinates for the description of \eqref{ns5} and \eqref{lst}. This is the task for the rest of this section. Before we start, we shall write \eqref{ns5} and \eqref{lst} generically as
\begin{align}
\label{generic}
ds^2&=-f_1(r) dx_1^2+\sum_{j=2}^6 dx_j^2+
A_\mathrm{i}(r)\left(\frac{dr^2}{f_1(r)}+r^2 d\Omega_3^2\right)\,,
\end{align}
with 
\begin{equation}
f_1(u)=1-\frac{u_0^2}{u^2}\,,
\end{equation}
and with $A_{\mathrm{i}}(r)$ chosen  properly  in each case.

\subsection{Kruskal coordinates}

 As is customary we shall write the line element \eqref{ns5} in terms of the $U(x_1,r)$ and $V(x_1,r)$  Kruskal coordinates\footnote{The parameter $a$ is introduced to make the arguments in the functions
 dimensionless.}
 \begin{equation}
\label{UV}
U=-e^{c_\mathrm{i} (F_\mathrm{i}(x)-a x_1)}\,,\quad V=e^{c_\mathrm{i} (F_\mathrm{i}(x)+a x_1)}\,,
\end{equation}
where $c_i$\footnote{The subindex $i$ denotes the model \eqref{ns5} or \eqref{lst}.} is a constant chosen on a case-by-case basis, see below, and $F(x)$ refers to the Tortoise coordinate in \eqref{generic}
\begin{equation}
\label{FF}
F_{\mathrm{i}}(x) =
\int dx\,\frac{\sqrt{A_{\mathrm{i}}(x)}}{f_1(x)}\,.
\end{equation}
After using \eqref{UV} one obtains for \eqref{generic} \cite{Pons2023}
\begin{equation}
ds^2 =-\frac{{\rm e}^{-2 c_\mathrm{i} F_\mathrm{i}(a\, r)}\,|f_1(a\, r)|}{a^2\, c_\mathrm{i}^2}dU\, dV +\sum_{j=2}^6 dx_j^2+
A_{\mathrm{i}}(r)\left( r^2 d\Omega_3^2\right)= - \Omega^2(r) dU\, dV+ \ldots \,,
\label{conformal}
 \end{equation}
 where the arguments inside the functions emphasise that the relations \eqref{UV} cannot be analytically inverted in most situations. Both solutions, \eqref{ns5} and \eqref{lst}, have a $\mathbb{S}^3$ term which plays a trivial role and we are free to ignore these directions for the time being\footnote{This is equivalent to considering only s-wave emission.}. 
Note that at this stage the prefactor $\Omega$ in \eqref{conformal} vanishes or becomes infinite at the event horizon, except for a specific value of the constant $c_i$, which is dictated precisely so that  $\Omega$ does not vanish in the entire patch. The results for $\cc{LST}$ and $\cc{NS5}$ are given in the equations below.
For both models we can perform the integral \eqref{FF} analytically, obtaining up to an arbitrary additive constant\footnote{Henceforth $y$ will be a dimensionless variable
\begin{equation}
\label{dimensionless}
    y=:r/r_0\,.
    \end{equation}
    }
\begin{enumerate}
\item [LST]
\begin{equation}
\label{F1}
F_{\mathrm{LST}}(y)=\frac{b}{2} \log (y^2-1)\,,\quad \cc{LST}=\frac{1}{b}\,,\quad
\Omega^{2}(y) =  b^2\frac{r_0^2}{ y^2} \,.
\end{equation}
And 
\item[NS5]
\begin{align}
\label{F2}
F_{\mathrm{NS5}}(y)= & \sqrt{y^2+b^2}+\frac{1}{2} \sqrt{1+b^2} \log\left|\frac{\sqrt{1+b^2}-\sqrt{y^2+b^2}}{\sqrt{1+b^2}+\sqrt{y^2+b^2}}\right|\,,\quad
\cc{NS5}=\frac{1}{\sqrt{b^2+1}}\,,\\ \nonumber
\Omega^{2}(y) &=  \frac{r_0^2}{ y^2} e^{-2\sqrt{\frac{b^2+y^2}{b^2+1}}} (1+b^2)^2 \left(1+\sqrt{\frac{b^2+y^2}{b^2+1}}\right)^2\,,
\end{align}
\end{enumerate}
where we have defined
\begin{equation}
\label{B}
b^2:=\frac{N}{m_s^2 r_0^2}\,.
\end{equation}
Note that $F(y)$ in \eqref{F1} and \eqref{F2} diverge as we approach the event horizon.
This is a consequence of the Kretschmann scalar of \eqref{ns5} or \eqref{lst} having a singularity at $r\rightarrow 0$.
Furthermore, $\Omega(y)$ is finite on the horizon or at large distances.  As a consistency check on our setting we have verified that in both cases $U\, V\rightarrow 0$, \eqref{UV}, at the event horizon.

\begin{figure}[h]
\begin{center}
\includegraphics[width=14cm,angle=0,clip=true]{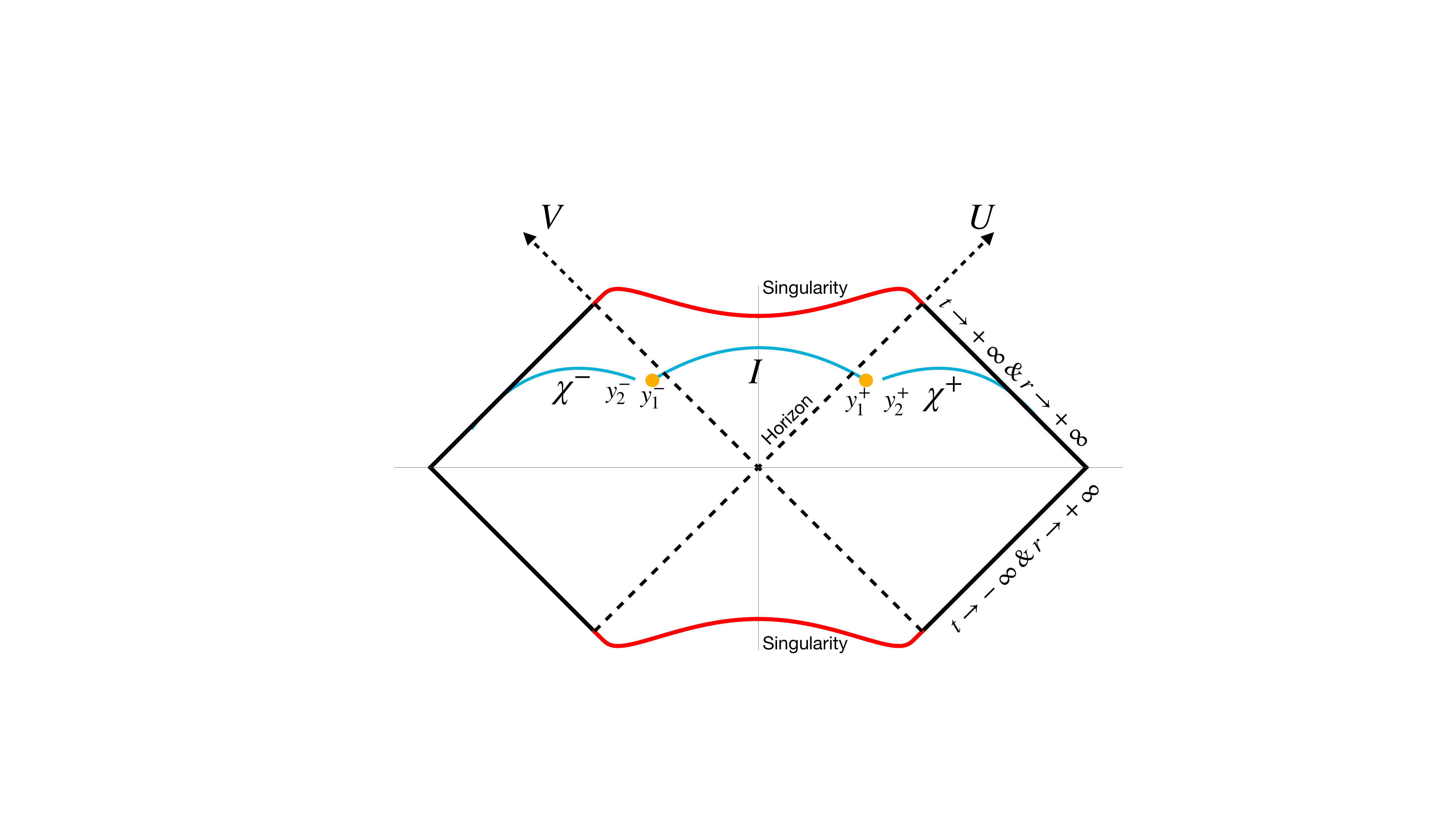}
\end{center}
\caption{\sl Conformal diagram for either \eqref{ns5} or \eqref{lst}. Each point in the diagram represents a two-dimensional Euclidean space, in our case extending in the directions $\{x_1,u\}$ of \eqref{generic}. Although the conformal factors (\ref{F1}, \ref{F2}) are different, their causal structures are the same. 
Red curves show the singularity at $r\to 0$. Dashed lines represent the horizon, $r\to r_0$.  Finally, solid black lines represent the
asymptotic. One salient point is that the $r\to 0$ surfaces bend the conformal diagram, as does the AdS Schwarzschild black hole \cite{Fidkowski:2003nf}.
  The region $\chi$, with states identified with the Haking radiation, is divided into two parts, $\chi^-$ and $\chi^+$. The boundary surfaces of $\chi^\pm$ are $y_2^\pm$ respectively. 
  We show the configuration containing an island, $\mathrm{I}$, that extends outside the horizon. Its boundaries are located at $y_1^\pm$.  
  The goal is to compute the entanglement entropy of the union $\mathrm{I}\cup \chi^+$. 
} 
\label{figu1}
\end{figure}

In Fig. \eqref{figu1} we represent the conformal diagram, $\{U,V\}$ coordinates \eqref{UV}, for \eqref{generic}.

\section{Entanglement Entropy}
\label{EELT}
We are going to follow the now well-known standard analysis in order to find the entanglement entropy. We take a two-sided black hole, as shown in Fig. \eqref{figu1}, which is initially in a pure state. Over time, it will become increasingly entangled with the thermal bath\footnote{It is evident that the Page curve we shall obtain is that of a non-gravitating theory which satisfies the {\sl split property of local QFT} \cite{raju} and {\sl not} that of the black hole.}. Thus its entropy is initially given by that of the matter entropy in the bulk while the black hole is there to provide the fixed curved background. At the Page time, the $\mathcal{O}(G_N^{-1})$ entanglement between the left and right black holes is replaced
by the  $\mathcal{O}(G_N^{-1})$ entanglement between the individual black holes and the bath \cite{Page:2013dx}. So there are at least two competing surfaces and the
generalised entropy is given by \cite{Almheiri:2019yqk}
\begin{equation}
\label{genentro}
    S(\chi)=\mathrm{min}_I\left[ \mathrm{ext}_I\left( \frac{{\mathcal{A}}(\partial I)}{4G_N}+S_{\rm matter}(\chi\cup I)\right) \right]\,.
\end{equation}
The first term of \eqref{genentro} corresponds to the area of the edge of the island contribution while the second is the von Neumann entropy computed with the quantum field theory formalism in the absence of gravity\footnote{See the final remark at the end of subsection \ref{subcbig}.}. This term depends on the relative location of the radiation with respect to the black hole horizon.

\subsection{Entanglement entropy for a single interval}

In a $2d$ CFT there is a universal expression for the entanglement entropy of an interval of length $L$, where the edges are very far apart \cite{Calabrese:2004eu}. This expression, in turn, coincides with the matter entropy per unit area of the metric \eqref{conformal} if we ignore the role of the $\mathbb{S}^3$ \cite{Fiola:1994ir}\footnote{See fig. \eqref{figu1} for notation.},
\begin{equation}
\label{EE1}
S^b_{\rm matter} 
=\frac{c}{3} \log[\frac{L}{ \epsilon}]
\approx
 \frac{c}{6} \log\left[\frac{1}{\epsilon^2}\frac{z_{y^+y^-}^2}{\Omega(y^+)\Omega(y^-)}\right]
\end{equation}
where $\epsilon$ is a uv regulator, $c$ is the central charge and
\begin{equation}
    z^2_{ab}=\Omega^2(a)\Omega^2(b) \left[U(b)-U(a)\right]\left[V(b)-V(a) \right]\,
    \end{equation}
is the geodesic distance between two boundary points in Kruskal coordinates
with $y^+=(t_y,y)$ and $y^-=(-t_y+i\frac{\pi}{2},y)$\footnote{By this convention, the signs of $U$ and $V$ in the left wedge of the Kruskal diagram picks and extra minus sign.}. Although \eqref{EE1} has been found in $2d$, there are actually
several numerical checks on its consistency, under some restrictions, in higher
dimensional space-times \cite{Almheiri:2019psy}.
The expression \eqref{EE1}
contains both ultraviolet and infrared divergences. The former are taken into account
with the standard renormalization technique and the overall outcome is the renormalization of the Newton's constant $G_N$. The latter divergences arise when one of the $y^\pm$ approaches the horizon.
Due to the general change of coordinates \eqref{UV}, \eqref{EE1} can be written as
\begin{equation}
\label{Sgen}
   S^b_{\rm matter} 
\approx \frac{c}{3} \log\left[ e^{\cc{i} F_{\mathrm{i}}(y)} {\rm cosh}
\left( \cc{i} \frac{t_y}{r_0} \right) \right] +
\frac{ c}{3}  \log\left[2\, \Omega(y) \right] \,.
\end{equation}
The details of the setup, the form of $F(x)$ and $\Omega(x)$, are not important at this point.
With \eqref{Sgen} at face, it is not surprising that, with $\cc{i}$ as a constant, all the gravitational models will behave linearly at large time, once we take into account the proper subtractions. In this regime, and provided that \eqref{EE1} is valid only for spacetime points far from the horizon, the remaining  terms in \eqref{Sgen} which do not
 directly involve time explicitly, are subleading.

We pause to motivate the subsequent steps in our study before proceeding with the standard analysis, see for example \cite{Hashimoto:2020cas}. 
Note that the relevant quantity within the argument of the time-dependent part of \eqref{Sgen} is $\cc{i}\,t$. In most of the cases studied in the literature, the  constant $\cc{i}$ is a harmless numerical factor\footnote{We have checked this claim in several models either with single or multi-horizons.} but due to the presence of the function $A_{\mathrm{i}}(r)$ in \eqref{conformal} it now plays a crucial role, since it introduces the free parameters that define the theory.
Thus, in the sequel, and in contrast to previous studies, it is not an early/late epoch
expansion what characterises the study of the black hole emission but the interplay of the factors in the $\cc{i}\,t$\ relation. More specifically,
 we are interested in the behaviour of the 
entanglement entropy with the dependence of $c_i$ on the number of branes $\cc{i}(N)$.

\subsubsection{The NS5 and LST entanglement entropies}

To make the point clearer,  let us have a closer look at the
possible values of $c_i(N)$ for the  \eqref{ns5} and \eqref{lst} models.
First of all, because they are a supergravity solution, the effective string
coupling must be bounded at its maximum value, the horizon. This leads to
the  constraint
\begin{equation}
    \label{limit1}
    N\ll m_s^2 r_0^2\,.
\end{equation}
Second, the scalar curvature should be small if the classical
geometry is to hold
\begin{equation}
\label{limit2}
    \mathcal{R}\sim \frac{1}{N}\,, \quad N\gg 1\,.
\end{equation}
Combining \eqref{limit1} and \eqref{limit2} we get
\begin{equation}
\label{limit3}
    1\ll N\ll m_s^2 r_0^2\,.
\end{equation}
With \eqref{limit3} at hand, we can conclude that although the two models are formally identical, see \eqref{generic}, the behaviour of the associated $c_i$ functions, \eqref{F1} and \eqref{F2}, is completely different at large-N
\begin{equation}
\label{limits}
    \cc{LST}\gg 1\,,\quad 
    \cc{NS5}\rightarrow 1\,.
\end{equation}
Bearing in mind that approximation the late time entanglement entropies \eqref{Sgen} are
\begin{align}
\label{larget}
\mathrm{LST}\quad S(\chi)=S^b_{\rm matter}&\approx  \frac{ c}{3} \frac{1}{ \sqrt{N}} \frac{t_y}{l_s}\,, \\
\mathrm{NS5}\quad S(\chi)=S^b_{\rm matter}&\approx  \frac{c}{3} \frac{t_y}{r_0}\,.\label{larget2}
\end{align}
As expected, the entanglement entropy grows linearly with time in both models.
A few words are in order.
Roughly speaking, entropy, \eqref{thlst}, gives the amount by which a system, let us call it $\mathcal{A}$, interacts with itself. Whereas the entanglement entropy, \eqref{larget}, captures the interactions between two of its subsystems, $\mathcal{A}_1\,, \mathcal{A}_2 \in \mathcal{A}$. What \eqref{larget} shows is that
if you have a parametrically large number of NS5-branes, $\mathcal{O}(N)\gg \mathcal{O}(1)$, but with \eqref{limit3} still valid, the system becomes non-interacting, the Hilbert space factorises and therefore we cannot recover information from outside \cite{Lorente-Espin:2007jqj}. This is equivalent to saying that the system becomes {\it non-interacting} with its boundary where the radiation is collected. In addition, in this limit the system can be considered as consisting of non-interacting strings \cite{PhysRevD.40.2626,Lowe:1994nm}, where the string tension is related to the usual string tension as \cite{HARMARK2000285}.
\begin{equation}
    \hat{\tau}=\frac{\tau}{N}\,.
\end{equation}
Thus the Hagedorn string tension is quantized to a fractional unit of the ordinary sting tension and vanishes at large-N.

Just for completeness, we roughly estimate the so-called Page time: As time goes on, \eqref{larget} has some conflict with the von Neumann entropy finiteness because the entropy radiation becomes larger than that of the black hole, which has only a finite number of degrees of freedom. 
This happens at
\begin{equation}
\label{tns5}
 t_y\gtrapprox N m_s^2 r_0^2\,,
\end{equation}
which proves to be extraordinarily long for the \eqref{lst} model.

\subsection{Entaglement entropy for two-disjoint intervals}
 
Now it's time to turn to the {\emph{island contribution} and see if the previous pattern has been washed out or still remains.
We now look  for non-trivial configurations for the bath at late times. We consider the left$\leftrightarrow$right symmetric island AB in fig \ref{figu1}. Then the entanglement entropy is given by
\begin{equation}
    S(\chi)=\frac{{\mathcal{A}}(\partial I)}{4G_N}+S^a_{\rm matter}\,.
\end{equation}
 These islands can be located either inside or outside the horizon depending on the specific gravitational model. This is equivalent to look for the entanglement entropy of two disjoint intervals inside $\mathbb{R}^2$ \cite{Casini:2005rm} between the points $(y^+_1,y^-_1)$ and $(y^+_2,y^-_2)$ in vacuum state. The contribution of the matter can be factorised as follows
\begin{equation}
\label{m}
S^a_{\rm matter}= \frac{c}{3} \log\left[\frac{z_{y_1^+y_1^-} z_{y_2^+y_2^-} z_{y_1^+y_2^+} z_{y_1^-y_2^-}}{\epsilon^4 z_{y_1^+y_2^-} z_{y_1^-y_2^+}} \right]\,.
\end{equation} 
From a physical perspective \eqref{m} gives the entanglement entropy for the matter fields located between the radiation and the island regions. 

Using \eqref{UV} we again obtain the same functional dependence on time for both models, assuming $\cc{i}(N)$ as numerical constant,
\begin{align}
\label{smatt}
 S^a_\mathrm{matter}& \approx \frac{c}{3} \log\left[4\, e^{\cc{i}(N)(F({y_1})-F({y_2}))}{\rm cosh}\left(\cc{i}(N)\frac{t_{y_1}}{r_0}\right){\rm cosh}\left(\cc{i}(N)\frac{t_{y_2}}{r_0}\right)\right]  \\ & + 
\frac{c}{3} \log\left[
\frac{ \mathrm{ cosh} \left[\cc{i}(N) (F({y_1})-F({y_2}))\right]- \mathrm{ cosh}\left(\frac{c_\mathrm{i}(N)}{r_0}(t_{y_1}-t_{y_2}) \right) }{  \mathrm{ cosh} \left[\cc{i}(N) (F({y_1})-F({y_2}))\right] +\mathrm{ cosh}\left(\frac{c_\mathrm{i}(N)}{r_0}(t_{y_1}+t_{y_2}) \right)}
\right]+\frac{c}{3}\log[\Omega({y_1})\Omega({y_2})]\nonumber \,. 
\end{align}
The fact that \eqref{smatt} holds independently 
of any gravitational model, indicates that the radiation-bath coupling 
is somehow universal in all of them.

As we have already seen, \eqref{larget2}, the NS5-brane model follows the standard behaviour with respect to the $N$ dependence,
 and so from now on we shall concentrate mainly on the LST model.

The next step is to consider that the island is formed near the horizon\footnote{The $\epsilon$ parameter, $\epsilon\ll 1$, can be either positive or negative.}, $y_1\approx 1+\epsilon,$ and that $y_2$ is near the boundary, $y_2\gg 1$.
Then \eqref{smatt} becomes\footnote{The expression \eqref{X} coincides with the results of 
\cite{kehagias} once we set the function $\Omega=1$ and $\cc{LST}=1$.}
\begin{align}
\label{X}
S^a_\mathrm{matter}& = \frac{c}{6} \log\left[32 \epsilon  y_2^2 {\rm cosh^2}\left(\cc{LST}(N)\frac{t_{y_1}}{r_0}\right){\rm cosh^2}\left(\cc{LST}(N)\frac{t_{y_2}}{r_0}\right)\right]
  \nonumber \\ & + 
\frac{c}{3} \log\left[
\frac{ 1-2 \frac{\sqrt{2\epsilon}}{y_2}  \mathrm{ cosh}\left(\frac{c_\mathrm{LST}(N)}{r_0}(t_{y_1}-t_{y_2}) \right) }{  1+2 \frac{\sqrt{2\epsilon}}{y_2}  \mathrm{ cosh}\left(\frac{c_\mathrm{LST}(N)}{r_0}(t_{y_1}+t_{y_2}) \right)}
\right]
+\frac{c}{3}\log[\Omega({y_1})\Omega({y_2})]
\,. 
\end{align}
We shall now comment on the limiting behaviour, \eqref{limits}, of \eqref{X}. 

\subsubsection{$\mathcal{O}(\mathrm{\cc{LST}}(N)) \lesssim \mathcal{O}(t)$}
This first limit is the standard one. We shall consider a large, but finite, time evolution and also a $\mathrm{\cc{LST}}(N)$ at most of the same order as time with
the constraint
\begin{equation}
\label{approx}
 1\gg   2 \frac{\sqrt{2\epsilon}}{y_2} \mathrm{cosh}\left(\frac{\cc{LST}(N)}{r_0}(t_{y_1}-t_{y_2})\right)\,.
\end{equation}
With this approximation in mind and a little of algebra, we get
\begin{equation}
S^a_\mathrm{matter}=    \frac{2}{3} c\log y_2 -\frac{2}{3} c \frac{\sqrt{2\epsilon}}{y_2}
{\rm cosh}\left(\frac{\cc{LST}(N)}{r_0}(t_{y_1}-t_{y_2})\right)+\frac{1}{3} c\log\left[\Omega(1+\epsilon) \Omega(y_2)\right]\,,
\end{equation}
which has an extreme for the values
\begin{equation}
\label{extremal}
    t_{y_1}=t_{y_2}\,,\quad \epsilon = \frac{1}{2 y_2^2}\,.
\end{equation}
We notice that, firstly, the parameter $\epsilon$ decreases as $y_2$ approaches the boundary. And secondly, because $\epsilon >0$ and assuming no further interaction the edge of the \emph{island} extends outside the horizon, see fig. \ref{figu1}. 
Because of the values \eqref{extremal}, the entanglement entropy \eqref{genentro} is
\begin{equation}
\label{sisland}
    S(\chi)\approx \frac{{\mathcal{A}}(\partial I)}{4G_N}\,.
\end{equation}


\subsubsection{$\mathcal{O}(\mathrm{\cc{LST}}(N)) \gg \mathcal{O}(t)$}
\label{subcbig}

As in the previous case, we take time to be large but bounded. Note that in this second case we can choose $\cc{LST}(N)$ as large
as we like. If we assume $t_a\ne t_b$ this inverts the inequality in \eqref{approx}.
This second limit is equivalent to considering a small time expansion in \eqref{X}. The result is
\begin{align}
    S^a_\mathrm{matter}& = \frac{c}{6} \log\left[32 \epsilon  y_2^2 {\rm cosh^2}\left(\cc{LST}(N)\frac{t_{y_1}}{r_0}\right){\rm cosh^2}\left(\cc{LST}(N)\frac{t_{y_2}}{r_0}\right)\right]
  \nonumber \\ & -
\frac{4}{3} c \frac{\sqrt{2\epsilon}}{y_2}  \mathrm{ cosh}\left(\frac{c_\mathrm{LST}(N)}{r_0} t_{y_1} \right)   \mathrm{ cosh}\left(\frac{c_\mathrm{LST}(N)}{r_0}t_{y_2} \right)
+\frac{c}{3}\log[\Omega(1+\epsilon)\Omega({y_2})]
\,,
\end{align}
which has no extrema  in the real domain. This indicates that, in this limit, 
the entanglement entropy of a single interval, \eqref{larget}, dominates the entire evaporation process  and, as we have seen, the latter decreases monotonically for higher value of N.

\section{Invariance}
\label{Inv}

There is one final question. Since the coordinates in \eqref{generic} have no physical meaning except to label space-time directions, one may wonder whether a coordinate redefinition that includes the N-factor within $A_\mathrm{LST}$ can bring \eqref{larget} into the form \eqref{larget2}. Inspection of \eqref{larget2} shows that the ratio 
$\frac{\mathrm{time}}{\mathrm{space}}$ must be invariant under scaling. This fixes $\mathrm{scaling(time)}=\mathrm{scaling(space)}$. This scaling should be done in such a way that \eqref{lst} is still a supergravity solution, which is a bit more involved that 
the simple Schwarzchild case \cite{doi:10.1142/S0217751X11053924}. In particular we look for pulling out all N-dependence in the action
\begin{equation}
    \label{I}
 {\cal I}={\cal F}(N)\left({\cal I}_{\rm grav}
+{\cal I}_{\rm surf}\right)\,. 
\end{equation}
The former is the Einstein-Hilbert
action
\begin{equation}
\label{gravaction}
  {\cal I}_{\rm grav}= {1\over  2
\kappa^2_{10} } \int_{\cal M} d^{10}x \sqrt{g} \left( R-{1\over 2}
\partial_\mu \phi \partial^\mu \phi  -{1\over 12} e^{-\phi}
 H_{(3)}^2\right)\,,
\end{equation}
while the latter is the surface contribution
\begin{equation}
    \label{surf} {\cal I}_{\rm surf}= {1\over  \kappa^2_{10} }
\oint_\Sigma  K d\Sigma\,, 
\end{equation}
with ${\cal M}$ being a ten-volume
enclosed by a nine-boundary $\Sigma$. The surface contribution is
determined by the extrinsic curvature
 \begin{equation}
     \label{K}
 K= \nabla_\mu n^\mu = {1\over\sqrt{g}} \partial_\mu \left( \sqrt{g}\, n^\mu \right)\,,
 \end{equation}
  where $n^\mu$~is the boundary outward normal vector.
 Finally one finds a constant NS-NS three-form along the $S^3$,
$H_{(3)} = 2 N \epsilon_3$. 

Note, that the rescaling of the coordinates simply reads as a relation between the scalar curvature of \eqref{lst} and the new metric tensor, which must not contain explicit 
N factors. Since the function $f_1$ is scale independent the first and last terms in \eqref{lst} specify that $x_1$ and $u$ must have the same scale dependence. This choice is inconsistent with the scaling of the function $A_\mathrm{LST}$, the third term in \eqref{lst}.
This simple exercise shows that the behavior of \eqref{larget}  is not an artifact of coordinatisation.

\section{Some Remarks}
\label{FR}

Let us summarise our findings. In the early stages of the radiation emission, we can prevent the NS5-brane system 
from interacting with its environment by setting a large-N limit, $\mathcal{O}(t) \ll \mathcal{O}(\sqrt{N})$. In this limit, neither
the string components of the system interact with each other,  nor does the system interacts 
with its surroundings, \eqref{larget}. In fact, the statement is valid up to time
scales around $\mathcal{O}(t)\sim \mathcal{O}(\sqrt{\frac{m_s^2 r_0^2}{N}})$, see
conclusions of section \ref{subcbig}. This change in behavior from the usual one at large-N could be due to the alleged phase transition suggested in \cite{Atick:1988si} although we have not stronger arguments for this. 
Evidently, when N is just a very large number, the entanglement entropy is simply suppressed and it scales
linearly with time, as usual.  

For very large times, $\mathcal{O}(t)\gg \mathcal{O}(\sqrt{\frac{m_s^2 r_0^2}{N}})$, there is an entanglement between the system and the surroundings, \eqref{sisland}, and it is no longer suppressed in the large-N limit. Note that this configuration is at work much earlier than
 the Page time, \eqref{tns5}. 

As a final remark, note that the model \eqref{lst} does not contain massless gravitons in its spectrum, as the limit \eqref{limith} decouples them from gravity, but still at finite N there would be 
entanglement entropy. This reinforces the statement that the entropy \eqref{genentro} 
does not take gravitational effects into account, and that only radiation plays a role.

\section{Some generalizations}
\label{gen}

It is tempting to ascribe the behaviour of \eqref{larget} to the existence of a maximum temperature, but as we shall
see, this behaviour seems to be inherent to the large-N limit and is most likely common to a larger family of 
gravitational backgrounds, namely those that are not scale invariant as described in section \ref{Inv}. Below we show
one of such background.

The metric \eqref{lst} is the ultraviolet completion of a large family group of regular non-
abelian monopole solutions in ${\cal N} = 4$ gauged supergravity, interpreted as 5-branes wrapped
on a shrinking $S^2$ \cite{Gubser:2001eg}. In the following we shall deal with a thermal deformation of one of
such metrics dual to ${\cal N} = 1$ SQCD with a superpotential coupled to adjoint matter \cite{Casero:2006pt}.

The metric field in Einstein frame\footnote{In accordance with  \cite{kehagias} we find that the value for the entanglement entropy 
is independent of the frame we have used.} is given by
\begin{eqnarray}
ds^2 =  e^{\phi_0/ 2} r &&\left\{ - K(r) dx_1^2 + \sum_{j=2}^4 dx_j^2 
+N \alpha^\prime \left( \frac{4}{ r^2 K(r)} dr^2 + \frac{1}{ \xi} d\Omega_2^2 
+\frac{1}{ 4-\xi} d\tilde{\Omega}_2^2 \right) \right. \nonumber \\
&&\left.  + \frac{N \alpha^\prime}{ 4} \left( d\psi + \cos\theta d\varphi+
\cos\tilde{\theta}d\tilde{\varphi}\right)^2 \right\}\,,
\quad K(r) = 1-\left(\frac{r_0}{ r}\right)^4\,.
\end{eqnarray}
In addition we have a dilaton field which is linear, $\phi = \phi_0+r\,,$~ and a RR 3-form field.

For this model  the expression for \eqref{FF} is almost identical to that obtained in the LST model \eqref{F1} 
\begin{equation}
\label{F3}
F_{\mathrm{}}(y)=\frac{b}{2} \log (y^4-1)\,,\quad c=\frac{1}{b}\,,\quad
\Omega^{2}(y) =  b^2\frac{r_0^2}{ y^3} \,.
\end{equation}

After a bit of algebra one recover the suppressed behaviour at large-N, \eqref{larget} 
\begin{align}
 S(\chi)=S^b_{\rm matter}&\approx  \frac{ c}{3} \frac{1}{ \sqrt{N}} \frac{t_y}{l_s}\,, \\
\end{align}
while, as expected, the corrections do not follow this pattern, as in \eqref{sisland}.

So the main message to take away is that some gravitational models may not be invariant under coordinate
reparametrisation. This fact introduces a parametric dependence in the thermodynamic quantities, which can lead to quantities that are  
suppressed in a particular corner of the parameter space.

\vspace{2cm}

{\bf Acknowledgements}

\noindent 
I thanks J. M. Pons for many useful discussions, especially for his valuable comments on the Kruskal coordinates.

This work is partially supported by grant PID2022-136224NB-C22, funded by MCIN/AEI/ 10.13039/501100011033/FEDER, UE.

\bibliographystyle{unsrt}
\bibliography{whitehole} 

\end{document}